\documentclass[aip,jcp,reprint,amsmath,amssymb,floatfix,citeautoscript]{revtex4-1}
\usepackage{color}
\usepackage{graphicx}
\usepackage{subcaption}
\usepackage{multirow}
\usepackage{threeparttable,booktabs}

\usepackage{ragged2e}
\usepackage[labelsep=period,format=plain,justification=justified,font=footnotesize]{caption}
\DeclareCaptionJustification{myjust}{\justifying}
\usepackage{txfonts}

\graphicspath{{image/}}
\allowdisplaybreaks

\begin{document}

\title{Co-iterative augmented Hessian method for orbital optimization}
\author{Qiming Sun}
\affiliation{Division of Chemistry and Chemical Engineering, 
California Institute of Technology, Pasadena, California 91125, USA}
\email{osirpt.sun@gmail.com}

\begin{abstract}
Orbital optimization procedure is widely called in electronic structure
simulation.
We developed a second order orbital optimization algorithm co-iteration
augmented Hessian (CIAH) method to search the orbital optimization solution.
In this algorithm, the orbital optimization is embedded in the diagonalization
procedure of the augmented Hessian (AH) equation.  Approximations to Hessian
matrix can
be easily applied in this method to reduce the computational costs.
We numerically tested the CIAH algorithm with the SCF convergence problem and
the Boys localization.  We found that CIAH algorithm has better SCF
convergence and less computational costs than direct inversion iterative
subspace (DIIS) algorithm.  The numerical tests suggest that CIAH is a stable,
reliable and efficient algorithm for orbital optimization problem.
\end{abstract}

\maketitle

\section{Introduction}

Self-consistency field (SCF) is the cornerstone of electronic structure
simulation.
It is required as a start point by almost all electronic structure calculations.
However, converging SCF to a reasonable solution is not a trivial problem.
Direct inversion iterative subspace\cite{Pulay1980,Pulay1982} (DIIS)
was the most successful method to accelerate the SCF convergence.
It was widely used in quantum chemistry program as the default optimization
algorithm for SCF problem.

Convergence problems are often observed for DIIS algorithm when the system has
open shell character or small HOMO-LUMO gap.
In the past, many SCF optimization techniques such as damping\cite{Cances2000}, level shift\cite{Saunders1973} and
enhanced DIIS algorithms EDIIS\cite{Kudin2002}, ADIIS\cite{Hu2010} were proposed to improve the DIIS
convergence\cite{Cances2001,Thoegersen2004,Thoegersen2005,Hoest2008,Host2008,Wang2011,Chen2011,Daniels2000}.
Although these techniques improved DIIS convergence performance,
DIIS and the improved algorithms have three main issues:
(i) As an error-vector based minimization method, DIIS algorithm does not guarantee the
SCF solution being the true minimum.  It is easy to have DIIS solution stuck at
saddle point;
(ii) DIIS algorithm does not honor the initial guess well.  The optimization
procedure may lead the wavefunction anywhere in the variational space;
(iii) DIIS algorithm does not have effective options to control the
optimization procedure.  Although DIIS convergence procedure can be
controlled by tuning the damping, level shift, subspace size,
extrapolation/interpolation constraints and other advanced
techniques\cite{Kudin2002,Hu2010,Cances2001,Thoegersen2004,Thoegersen2005,Hoest2008,Host2008,Wang2011,Chen2011,Daniels2000},
the influence are unpredictable.

The issues of DIIS algorithm can be surmounted in the second order SCF
optimization algorithm\cite{Fischer1992,Neese2000}.
Second order algorithm (Newton or quasi-Newton methods) directly
minimizes the function gradients with the assistance of Hessian matrix.
The Hessian matrix can provide an optimal displacement in the parameter
space and a judgement whether a solution is at saddle point or local minimum.
By tuning step size and trust region, one can easily control the optimization
procedure and constrain the solution to certain region.

In Newton's methods, explicitly constructing the Hessian matrix and its
inverse matrix often leads to high computational cost.
Quasi-Newton methods address the cost problem by approximating the Hessian
matrix (or its inverse) based on the change of gradients during the optimization.
Such Hessian approximation may be a native choice for complicated object function.
This is not an optimal scheme for SCF problem because the simple structure of
SCF Hessian matrix is not recognized by the gradients-oriented quasi-Newton methods.
It is not necessary to estimate the Hessian based on the updates of gradients.
Stripping Hessian evaluation from gradients construction, one would gain
larger flexibility to approximate the Hessian matrix.
Since Hessian matrix serves mainly as an auxiliary metric to adjust the
descending direction of the displacement,
the accuracy of Hessian matrix is not critical to convergence procedure.
Therefore, many physical significant considerations rather than pure numerical
treatment can be brought into the Hessian approximation.

A practical problem for Newton's method is the
treatment of the negative eigenvalue and the singularity of the Hessian matrix.
Negative Hessian is common when system is out of the quadratic region.
Non-invertible Hessian matrix is also regularly found near the saddle point.
These problems may result in wrong direction or singular displacement in the parameter space.
Level shift for Hessian is often used to fix the singularity problem and
adjust the descending direction.
Here, the augmented Hessian (AH) method\cite{Lengsfield1980,Jorgensen1983,Jensen1984},
which can be traced back to the early work in the mutli-configuration
self-consistent field (MCSCF) optimization, provided a decent solution to
dynamically adjust the level shift of Hessian eigenvalues.
In the state far from the quadratic region, AH works close to the gradient
descent method.  When the gradients approach to zero, AH equation turns to
the normal Newton's equation.


Based on the AH method, we present in this paper a second order optimization algorithm,
co-iterative augmented Hessian (CIAH) method for orbital optimization problem.
The basic idea of CIAH algorithm is to 
embed the optimization procedure into the diagonalization procedure of AH equation.
The structure of orbital optimization parameters are considered in the algorithm.
Particularly, small step is preferred by this algorithm than one shot
``optimal'' step.
The details are described in Section \ref{sec:algorithm}.
This algorithm is universal for a wide range of orbital optimization problem,
from Hartree-Fock and Kohn-Sham energy minimization for restricted and
unrestricted, closed and open shell, molecule and crystal, to MCSCF
optimization and orbital localization.
Since MCSCF energy minimization is beyond the pure orbital optimization, the
relevant algorithm details and convergence performance are documented in
our MCSCF work\cite{CASSCF}.
In Section \ref{sec:numeric}, we numerically verified the performance of the
algorithm with various kinds of SCF calculations and orbital localization.

\section{Algorithm}
\label{sec:algorithm}
Provided $E$ the energy functional of the one-particle orbital rotation $U$
subjecting to the exponential ansatz of the unitary transformation
\begin{gather}
  U = e^{\hat{R}}, \\
  \label{eq:u}
  \hat{R} = \sum_{pq} R_{pq} a_p^\dagger a_q, \\
  R_{pq} = -R_{qp}^*,
\end{gather}
energy minimization can be treated as a non-linear search problem for the
optimized $\mathbf{R}^*$ where the stationary condition holds
\begin{equation*}
  \frac{\partial E}{\partial R_{pq}}\Big|_{\mathbf{R}^*} = 0.
\end{equation*}
For $k$th iteration, a displacement $\mathbf{x}^{(k+1)}$
to approach the solution $\mathbf{R}^*$ can be obtained by solving the AH
matrix equation
\begin{gather}
  \begin{pmatrix}
    0 & \mathbf{g}^{(k)\dagger} \\
    \mathbf{g}^{(k)} & \mathbf{H}^{(k)}
  \end{pmatrix}
  \begin{pmatrix}
    1 \\
    \lambda\mathbf{x}^{(k+1)}
  \end{pmatrix}
  = \epsilon
  \begin{pmatrix}
    1 \\
    \lambda\mathbf{x}^{(k+1)}
  \end{pmatrix},
  \label{eq:aheig}
  \\
  g_{pq}^{(k)} = \frac{\partial E}{\partial R_{pq}}\Big|_{\mathbf{R}^{(k)}}
  \notag\\
  H_{pq,rs}^{(k)} = \frac{\partial^2 E}{\partial R_{pq}\partial R_{rs}}
  \Big|_{\mathbf{R}^{(k)}}.
  \notag
\end{gather}
The AH matrix here plays the role to damp the solution of Newton's method
\begin{equation}
  \mathbf{g}^{(k)} + \lambda (\mathbf{H}^{(k)}-\epsilon)\mathbf{x}^{(k+1)} = 0
  \label{eq:levelshift}
\end{equation}
with level shift
\begin{equation*}
  \epsilon = \lambda\mathbf{g}^{(k)\dagger} \mathbf{x}^{(k+1)}.
\end{equation*}
The level shift parameter circumvents the descending direction problem when the
optimizer is around the non-quadratic region.  When the answer approaches the
local minimum, Eq. \eqref{eq:levelshift} turns to the standard Newton's
equation because $\epsilon$ rapidly decays to zero.
The scaling factor $\lambda$ is commonly used in AH algorithm to adjust the
step size\cite{Jensen1984,Ghosh2008}.


In the CIAH program, we don't have sophisticated step size adjustment.
A special feature of orbital optimization is that the matrix elements
of the unitary transformation \eqref{eq:u} must lie in the range $[-1,1]$.
This allows us to fill the optimal rotation matrix
with a series of small displacements.
Therefore, we simply removed the $\lambda$ parameter in Eq \eqref{eq:aheig}
and scaled down the largest element of the displacement vector $\mathbf{x}$ to
a small predefined threshold $\delta$
\begin{equation}
  \begin{cases}
    \mathbf{x}^{(k)} & \max(\mathbf{x}^{(k)}) < \delta \\
    \frac{\delta}{\max(\mathbf{x}^{(k)})} \mathbf{x}^{(k)} & \text{otherwise}
  \end{cases}.
  \label{eq:scalestep}
\end{equation}
The thresholds are slightly different in different optimization problems.
For SCF and MCSCF wavefunction, we prefer smaller step size $\delta=0.03$ to
provide a smooth convergence procedure because we usually have reasonable
initial guess.
In the orbital localization problem, the initial guess is often very different
to the final answer.  Optimization often starts with canonical orbitals
and ends up with local orbitals.
A slightly larger threshold $\delta=0.05$ is used so that the optimizer can move
quickly in the parameter space.
Nonetheless, it should be noted that 
small step generally has advantage over large step in CIAH algorithm.

Because of the small step strategy, it can be expected that the Hessian matrix in
the adjacent iterations should be close to each other.
Approximately, one can keep the Hessian matrix unchanged and update only the
gradients during the optimization iterations.
This requires small modification to the Davidson diagonalization
program\cite{Davidson1975} which is used to solve the AH equation
\eqref{eq:levelshift}.
In the conventional AH algorithm, one expands and diagonalizes the
$(n+1)$-rank AH matrix in the subspace representation.
In the modified version, we only keep track of the subspace corresponding to
the $n$-rank Hessian matrix.
When the system moved to the new point,
we construct the gradients representation with the old $n$-rank basis
$\{v\}$ and obtain the new representation of AH matrix
\begin{align*}
  H_{ij}^{(k+1)} &= \langle v^{(k)}_i|H[\mathbf{R}^{(s)}]| v^{(k)}_j\rangle, \quad s < k, \\
  g_{i}^{(k+1)} &= \langle v^{(k)}_i|g[\mathbf{R}^{(k+1)}]\rangle.
\end{align*}

Besides the Hessian reservation treatment, we embedded the function
optimization iteration into the Davidson diagonalization iteration.
Within each cycle of CIAH updating, the AH diagonalization program
enlarges the subspace by one basis vector $\mathbf{v}$ in terms of the
Davidson preconditioner
\begin{equation}
  \mathbf{v}^{(k+1)} = (\mathbf{H}_0-\epsilon)^{-1}(\mathbf{H}\mathbf{x}^{(k)}-\epsilon \mathbf{x}^{(k)}).
  \label{eq:davidsonbasis}
\end{equation}
The Hessian matrix representation is therefore improved gradually during
the CIAH optimization cycles.
Due to the error in the diagonalization solver, the displacement vector
$\mathbf{x}$ might not be optimal in the early stage of the optimization.
The error can be removed in the later steps
and the optimal displacement will be generated when the AH diagonalization
solver gets enough bases to accurately represent the gradients and the Hessian
matrix.
The small step strategy plays an important role in the CIAH algorithm
because it reduces the negative effects of the poor displacement vector
appeared in the early optimization stage.

When the system is around the non-quadratic region, the main purpose of the orbital
Hessian is to adjust the direction of the displacement.  The accuracy of the
Hessian matrix is not highly important in this circumstance.
One can take coarse approximations for the orbital Hessian to reduce the
computational costs, eg
projecting the Hessian from low level basis sets,
or superposition of the fragment Hessian matrices.
Integral approximations such as density fitting,
high cutoff, 
sparse meshgrids (for DFT numeric integration),
or even single precision integrals can be used as well.

Unlike the Hessian approximations, it is less flexible to approximate the
orbital gradients.
Orbital gradients provide two aspects of usage:
the convergence criteria and a rough optimization direction.
The gradients must be accurately evaluated when it was used as the convergence
criteria.
For the optimization direction, approximated orbital gradients are acceptable.
The Tayler expansion of orbital gradients around given point $\mathbf{R}^{(k)}$ is
\begin{equation}
  g[\mathbf{R}] = g[\mathbf{R}^{(k)}] + H[\mathbf{R}^{(k)}] \cdot (\mathbf{R} - \mathbf{R}^{(k)}) + \dots
  \label{eq:approxgrad}
\end{equation}
If the new point $\mathbf{R}$ is close to the expansion point,
the gradients at $\mathbf{R}$ can be approximated by the first order expansion.
It should be noted that the first order approximation may cause large error
in orbital gradients, especially when approximated Hessian matrix is employed.
Our solution is to insert keyframes which are the exact gradients
in certain iterations (while the Hessian matrix is still approximated).
A keyframe is triggered by two conditions:
(i) if the gradients are out of trust region with respect to the previous keyframe,
(ii) if the number of iterations between two keyframes is more than the
predefined keyframe intervals.
The keyframe here provides not only the adjustment to the optimization path,
but also a reference to check whether convergence criteria are met.

Here we briefly summerized the CIAH algorithm (Table \ref{tab:ahiter} is an
example of the evolution of each quantities during the optimization procedure).
\begin{enumerate}
  \item \label{init}
Given an initial value $\mathbf{R}^{(0)}$, the optimizer starts to build the
AH equation with one trial vector and the AH matrix is a $2\times 2$ matrix.
Diagonalizing the AH matrix provides the first displacement $\mathbf{x}^{(0)}$
which is then scaled down to the predefined step-size threshold as shown by Eq.
\eqref{eq:scalestep}.
  \item \label{micro}
For $k$th iteration, the displacement $\mathbf{x}^{(k)}$ is used to update
the Davidson subspace basis $\mathbf{v}^{(k+1)}$ using Eq.
\eqref{eq:davidsonbasis} and the gradients $\mathbf{g}^{(k+1)}$ with the first
order approximation \eqref{eq:approxgrad}.
The new basis and the approximated gradients are used to build
the new AH matrix for the next displacement $\mathbf{x}^{(k+1)}$.
This step is applied many times unitil the number of iterations reaches
predefined upper limit (go to step \ref{macro}) or the keyframe is triggered
(go to step \ref{keyframe}).
  \item \label{keyframe}
In the keyframe ($\mathbf{g}^{(4)}$, $\mathbf{g}^{(7)}$ in Table
\ref{tab:ahiter}), gradients are evaluated exactly based on the cumulated
displacements $(\mathbf{R}^{(0)}+\mathbf{x}^{(1)} + \dots + \mathbf{x}^{(k)})$.
Based on the exact gradients,
there are three different conditions for the program flow: (i)
If the norm of gradients is smaller than the required convergence criterion,
we will call other convergence checks and prepare to finish the optimization;
(ii) if the gradients are very different to the last approximate gradients (out
of trust region, in which the ratio between the new and old gradients' norms is over 3),
the optimizer will move to step \ref{macro};
(iii) otherwise, we insert the exact gradients into the AH equation
\eqref{eq:aheig} to generate a better displacement vector then go back to
step \ref{micro}.
  \item \label{macro}
We move the system to the new point ($\mathbf{R}^{(1)}$ in Table
\ref{tab:ahiter}), then discard all bases of the Davidson diagonalization solver
and rebuild the AH matrix (as step \ref{init} did).
The last displacement is used as the initial guess of the Davidson
diagonalization solver.
The program will go back to step \ref{micro} and start a new cycle of
optimization.
We labelled such a cycle from step 2 - step 4 as a macro iteration.
\end{enumerate}

The number of iterations (micro iterations) within each macro iteration is
around 10 in our implementation.
In the SCF procedure, the matrix-vector produce of the Davidson
diagonalization algorithm are the most time-cosuming operations.
It involves the contraction of J/K (coulomb and exchange) matrices
which is as expensive as the regular Fock matrix construction.
The number of J/K contractions in CIAH algorithm is equal to the total number
of micro iterations (due to the matrix-vector operations in the AH matrix diagonalization)
plus the number of keyframes.
By using the Hessian approximations,
the real cost can be much lower than that the number of J/K contractions indicated
because the time-consuming J/K contractions are restricted in the keyframes.
In the CIAH algorithm, small step size is the guarantee for the Hessian
and the gradients approximations.
These approximations might lead to more macro iterations but generally it
reduces the cost of J/K operations thus improves the overall performance.

\begin{table*}
\centering
  \caption{The evolution of gradients and Hessians during the CIAH iterations}
\begin{tabular}{ccllcccccc}
  \hline
  Macro iter & AH subspace size & Hessians & Gradient & Displacement \\
  \hline
  1 & 1 & $\mathbf{H}^{(0)}=\frac{\partial^2 E}{\partial R\partial R'}[\mathbf{R}^{(0)}]$
          & $\mathbf{g}^{(0)} = \frac{\partial E}{\partial R}[\mathbf{R}^{(0)}]$  & $\mathbf{x}^{(1)}$ \\
    & 2 & $\mathbf{H}^{(0)}             $ & $\mathbf{g}^{(1)}\approx\mathbf{g}^{(0)}+\mathbf{H}^{(0)}\mathbf{x}^{(1)}$ & $\mathbf{x}^{(2)}$ \\
    & 3 & $\mathbf{H}^{(0)}             $ & $\mathbf{g}^{(2)}\approx\mathbf{g}^{(1)}+\mathbf{H}^{(0)}\mathbf{x}^{(2)}$ & $\mathbf{x}^{(3)}$ \\
    & 4 & $\mathbf{H}^{(0)}             $ & $\mathbf{g}^{(3)} = \frac{\partial E}{\partial R}
                  [(\mathbf{R}^{(0)}+\mathbf{x}^{(1)}+\dots+\mathbf{x}^{(3)})]$          & $\mathbf{x}^{(4)}$ \\
    & 5 & $\mathbf{H}^{(0)}             $ & $\mathbf{g}^{(4)}\approx\mathbf{g}^{(3)}+\mathbf{H}^{(0)}\mathbf{x}^{(4)}$ & $\mathbf{x}^{(5)}$ \\
    & 6 & $\mathbf{H}^{(0)}             $ & $\mathbf{g}^{(5)}\approx\mathbf{g}^{(4)}+\mathbf{H}^{(0)}\mathbf{x}^{(5)}$ & $\mathbf{x}^{(6)}$ \\
    & 7 & $\mathbf{H}^{(0)}             $ & $\mathbf{g}^{(6)} = \frac{\partial E}{\partial R}
                  [(\mathbf{R}^{(0)}+\mathbf{x}^{(1)}+\dots+\mathbf{x}^{(6)})]$          & $\mathbf{x}^{(7)}$ \\
    & 8 & $\mathbf{H}^{(0)}             $ & $\mathbf{g}^{(7)}\approx\mathbf{g}^{(6)}+\mathbf{H}^{(0)}\mathbf{x}^{(7)}$ & $\mathbf{x}^{(8)}$ \\
    \multicolumn{4}{l}{$\mathbf{R}^{(1)} = \mathbf{R}^{(0)} + \mathbf{x}^{(1)} + \dots + \mathbf{x}^{(8)}$} \\
  2 & 1 & $\mathbf{H}^{(1)}=\frac{\partial^2 E}{\partial R\partial R'}[\mathbf{R}^{(1)}]$
          & $\mathbf{g}^{(8)} = \frac{\partial E}{\partial R}[\mathbf{R}^{(1)}]$ & $\mathbf{x}^{(9)}$ \\
    & 2 & $\mathbf{H}^{(1)}             $ & $\mathbf{g}^{(9)}\approx\mathbf{g}^{(8)}+\mathbf{H}^{(1)}\mathbf{x}^{(9)}$ & $\mathbf{x}^{(10)}$ \\
  \hline
\end{tabular}
  \label{tab:ahiter}
\end{table*}

\section{Numerical assessment}
\label{sec:numeric}
The CIAH algorithm were implemented in the open-source electronic structure
program package PySCF\cite{PYSCF}.
In all SCF calculations, the convergence criteria are set to $10^{-10}$
$E_\text{h}$ for the change of energy and $10^{-5}$ for the norm of gradients.
For spatial and spin symmetry, symmetry broken is allowed if it can decrease the total energy.
The initial guess orbitals for CIAH solver are fed from the regular DIIS-SCF
iterations, of which the change of SCF energy converges to 1.0 $E_\text{h}$.
The DIIS-SCF calculations are initialized with the superposition of atomic
density. 
In the DIIS-SCF iterations, unless otherwise specified, DIIS subspace size is 8
and level shift is 0.2.  For ROHF methods, DIIS extrapolation is applied on
Roothaan's open-shell Fock matrix.  The structures of all molecules can be
found in the support material.

\subsection{Hessian approximations}

\begin{table*}
  \centering
  \caption{Number of macro iterations and key-frames and J/Ks required for
  different CIAH approximations in $^3$Cr$_2$ and $^3$Fe-porphine SCF
  optimization.}
  \label{tab:happrox}
  \begin{tabular}{llcccccccc}
    \hline
          &                 & \multicolumn{3}{c}{$^3$Cr$_2$} && \multicolumn{3}{c}{$^3$Fe-porphine} \\
          \cline{3-5} \cline{7-9}
          &                 & Macro iters & keyframes & J/Ks && Macro iters & keyframes & J/Ks \\
    \hline 
    ROHF  & Standard CIAH   & 18          & 51        & 304  && 3           & 8         & 30 \\
          & DF Hessians     & 19          & 51        & 308  && 3           & 8         & 30 \\
          & Projected basis & 18          & 37        & 184  && 14          & 24        & 63 \\
          & All in one      & 14          & 33        & 159  && 14          & 24        & 63 \\
    UB3LYP& Standard CIAH   & 17          & 38        & 150  && 9           & 21        & 84 \\
          & DF Hessians     & 17          & 39        & 170  && 10          & 21        & 87 \\
          & sparse grids    & 19          & 41        & 157  && 9           & 20        & 85 \\
          & DF +sparse grids& 17          & 39        & 153  && 9           & 20        & 85 \\
          & Projected basis & 26          & 49        & 183  && 22          & 39        & 147\\
          & All in one      & 25          & 47        & 164  && 22          & 39        & 147\\
    \hline
  \end{tabular}
\end{table*}

\begin{figure*}
  \begin{center}
\begin{subfigure}[b]{0.5\textwidth}
  \centering
  \includegraphics[width=\textwidth]{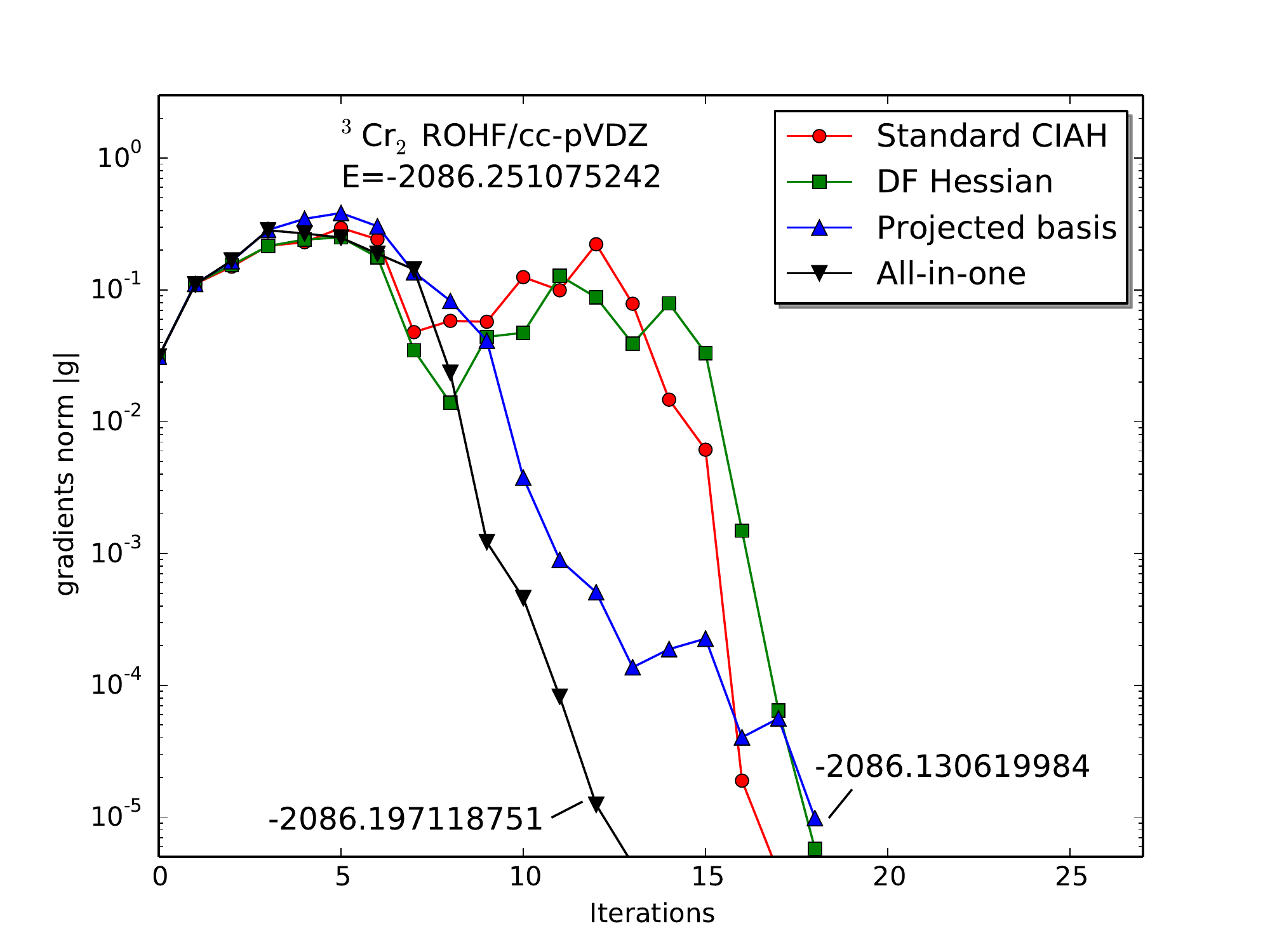}
  \label{fig:cr2:rohf}
\end{subfigure}%
\begin{subfigure}[b]{0.5\textwidth}
  \centering
  \includegraphics[width=\textwidth]{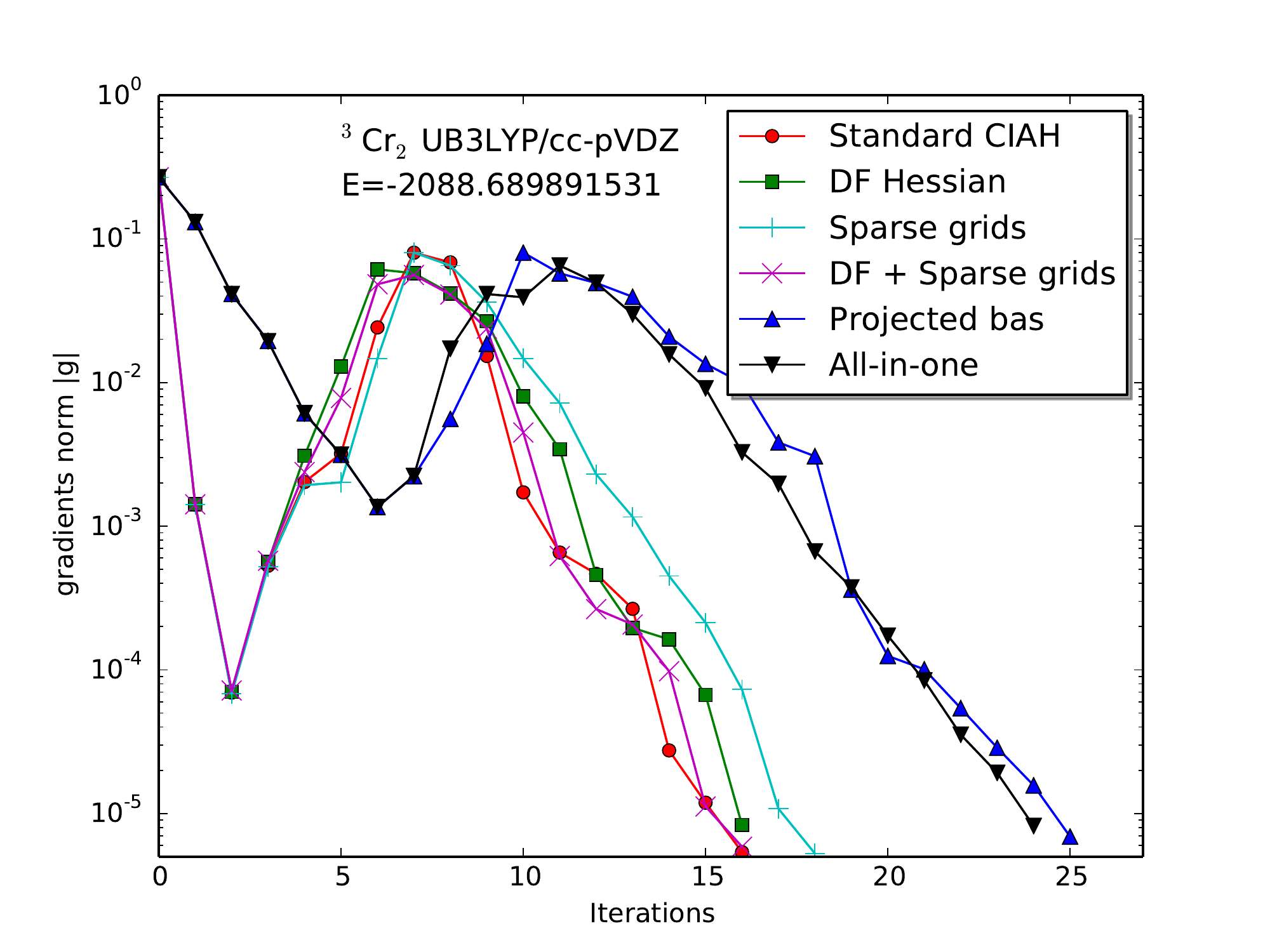}
  \label{fig:cr2:uks}
\end{subfigure}%
\\
\begin{subfigure}[b]{0.5\textwidth}
  \centering
  \includegraphics[width=\textwidth]{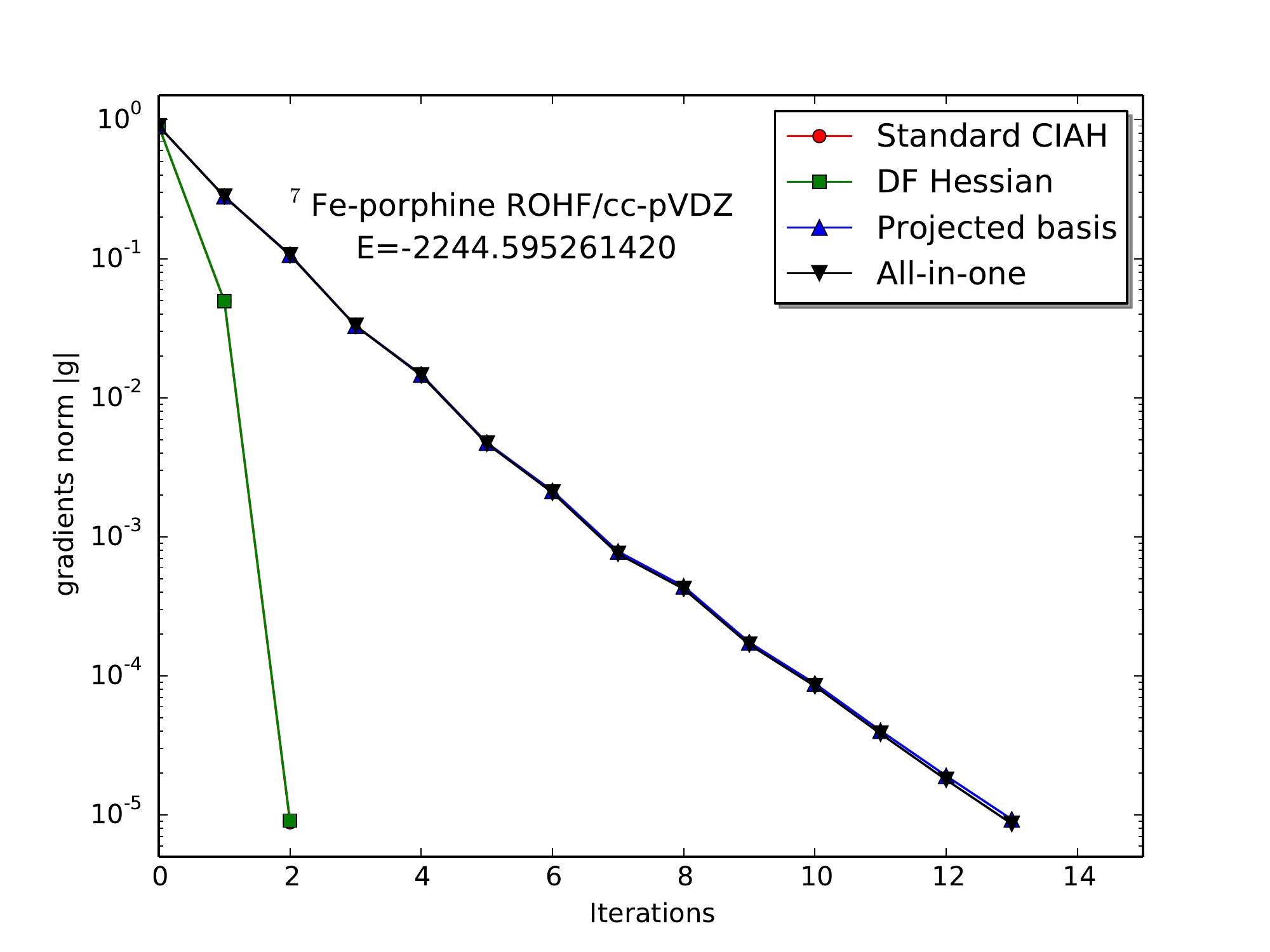}
  \label{fig:fepor:rohf}
\end{subfigure}%
\begin{subfigure}[b]{0.5\textwidth}
  \centering
  \includegraphics[width=\textwidth]{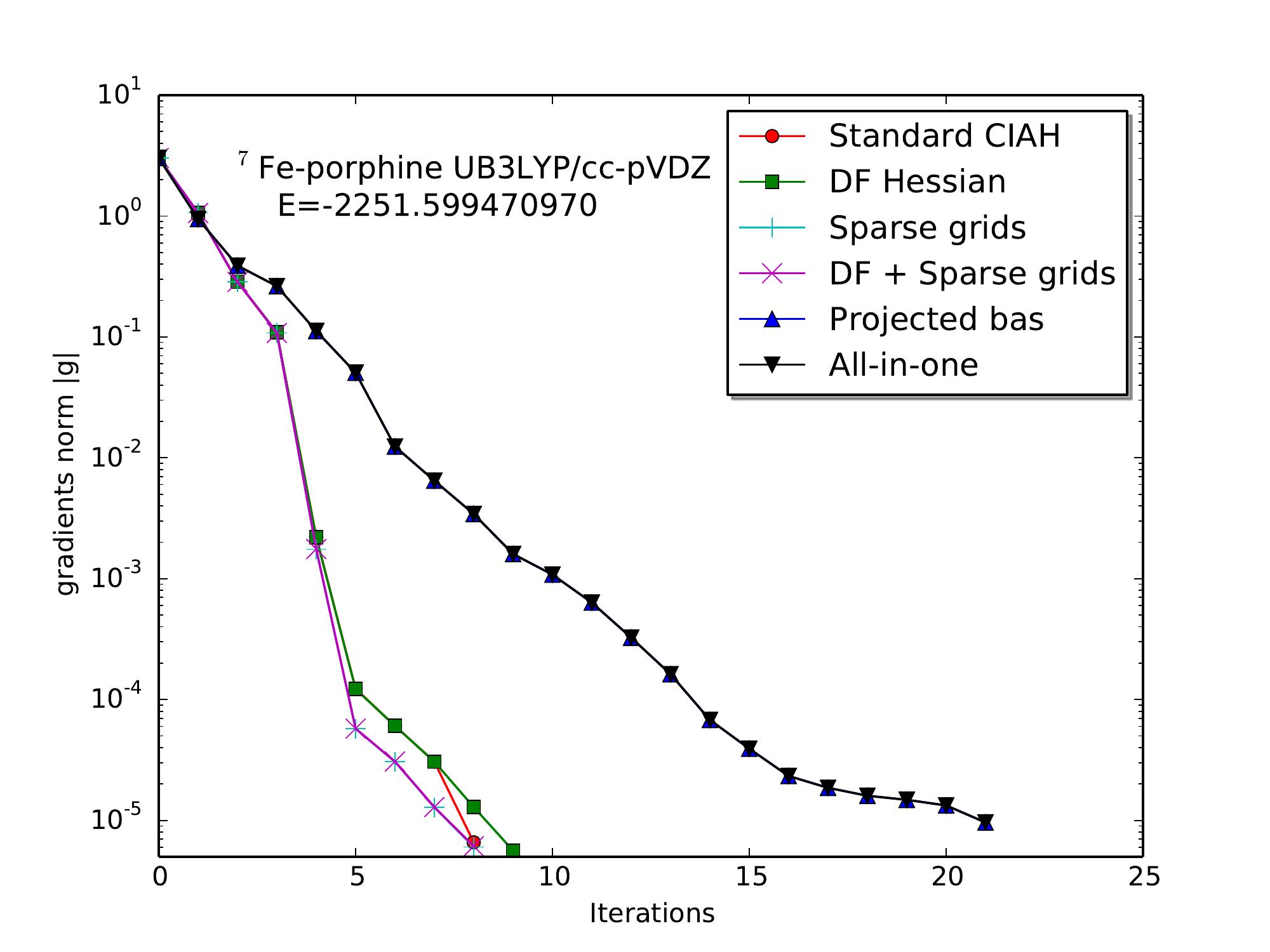}
  \label{fig:fepor:uks}
\end{subfigure}%
\end{center}
\caption{SCF convergence for different Hessian approximations.}
\label{fig:happrox}
\end{figure*}

We use the triplet states of Cr$_2$ and Fe-porphine with ROHF/cc-pVDZ
and UB3LYP/cc-pVDZ calculations to inspect the HF and DFT Hessian
approximations, including density-fitting (DF), basis projection, and
sparse grids (for DFT).
In the DF approximation, we employed Weigend Coulomb-fit basis as the auxiliary
fitting basis which is inaccurate for exchange integrals. 
In the sparse grids approximation, we computed the second order derivative of
XC functionals with small mesh grids in which the number of radial grids and
angular grids are (15,86) for light elements and (30,110) for transition metal
while the normal mesh grids are (75,302)/(90,434).
In the basis projection approximation, we expand the Hessian matrix on
single-zeta basis then transform back to cc-pVDZ basis.
The single-zeta basis is derived from cc-pVDZ basis by removing the outermost
one shell for each angular momentum.
The single-zeta basis significantly reduces the number of basis functions,
from 86 to 54 for Cr$_2$ and from 439 to 159 for Fe-porphine.
Level shift is not applied in the initial guess of $^3$Cr$_2$ because level
shift leads to a symmetry reserved solution which is high in total energy.

In Figure \ref{fig:happrox}, we compare the SCF convergence for standard CIAH
(without applying above approximations) and the approximate CIAH iterations.
The convergence curves of DF Hessian and sparse grids are close to the standard
CIAH curve in all tests, presenting that they have high quality approximations
to the Hessian matrix.
The basis projection approximation has large error because it misses large
fraction of the Hessian matrix.
Except the early stage of the optimization, such error leads to obvious
deviation to the standard CIAH convergence curve.
In three of the tests, this poor approximation can converge the SCF to the
right answer but require two times of the macro iterations or more.
For $^3$Cr$_2$ ROHF, there are several local minimum answers.
Depending on the numerical fluctuation during the optimization, basis
projection can occasionally converge to the lowest one.

Table \ref{tab:happrox} shows the number of macro iterations,
key-frames and J/K contractions for each Hessian approximations.
In the standard CIAH procedure, the costs to compute key-frame are roughly equal
to the costs of J/K contraction .
The total costs are determined by the number of J/K contractions which are
the dominant operations in the Hessian construction.
In the approximate schemes, the Hessian construction becomes less expensive.
The computational dominant step turns to the key-frame construction.
For schemes like DF and sparse grids, the number of key-frames is around
25\% of the number of J/K contractions.
One could get 2 - 3 times speed up for the overall performance.
For example, by using the density fitting for $^3$Fe-porphine UHF Hessian, the
Hessian construction takes only 25\% of the total computing time while it needs
over 70\% of the computing time in the standard CIAH treatment.
In practice, even the poor basis projection approximation has significant performance
advantage over the standard CIAH iteration since it requires much less
keyframes than the number of J/K contractions of the standard CIAH scheme.

\subsection{CIAH vs. DIIS}

Based on the tests of Hessian approximations,
the combination of DF and sparse grids provides an efficient scheme to
approximate the Hessian matrix without the penalty of the convergence rates.
They are adopted in our following SCF calculations.
Applying basis projection at early optimization stage can further improve the
total computing costs.
This treatment is not applied in the comparison because it is largely
associated with the initial guess than optimization iterations.

\begin{table*}
  \centering
\begin{threeparttable}
  \caption{CIAH vs. DIIS. Level shift 0.2 is applied in DIIS. 
  For the unconverged solutions, the values are reported up to (without) the
  oscillated decimal place.}
  \label{tab:alltests}
  \begin{tabular}{llllllllllllllllllllll}
    \hline
    Molecule & Method & CIAH\tnote{e} & $E$ & DIIS & $E$ \\
    \hline
$^1$Cr$_2$\tnote{a}
      &LSDA/3-21G     & 18 (40) & -2073.907478426 & 191     & -2073.907480839 \\
      &B3LYP/3-21G    & 5 (11)  & -2078.075421673 & 31      & -2078.075421674 \\
$^3$Cr$_2$\tnote{a}
      &U-LSDA/3-21G   & 55 (175)& -2073.949040592 & 328     & -2073.948413591 \\
      &U-B3LYP/3-21G  & 75 (135)& -2078.244163328 & 56      & -2078.175189832 \\
$^1$UF$_4$\tnote{a}
      &LSDA/LANL2DZ   & 15 (38) & -448.6767448225 & 184     & -448.6767448158 \\
      &B3LYP/LANL2DZ  & 23 (56) & -451.0320849544 & $> 500$ & -451.03129955   \\
$^3$UF$_4$\tnote{a}
      &ROHF/LANL2DZ   & 18 (53) & -448.7341401101 & 431     & -448.7176513294 \\
      &UHF/LANL2DZ    & 8 (27)  & -448.7364768642 & 424     & -448.7203771973 \\
      &U-LSDA/LANL2DZ & $>500$  & -448.7307       & 193     & -448.7310709215 \\
      &U-B3LYP/LANL2DZ&130 (188)& -451.0971319546 & 132     & -451.0970715829 \\
$^1$Ru$_4$CO\tnote{b}
      &RHF/LANL2DZ    & 13 (34) & -484.6768737985 & 125     & -484.6768611861 \\
      &B3LYP/LANL2DZ  & 8 (19)  & -488.4169663521 & 99      & -488.4169663513 \\
$^3$Ru$_4$CO\tnote{b}
      &ROHF/LANL2DZ   & 12 (30) & -484.7310915959 & $> 500$ & -484.675        \\
      &UHF/LANL2DZ    & 29 (82) & -484.9692302601 & $> 500$ & -484.6947       \\
      &U-B3LYP/LANL2DZ&123 (211)& -488.4555228944 & 280     & -488.4348633206 \\
$^3$Fe-porphine\tnote{c}
      &ROHF/cc-pVDZ   & 3 (8)   & -2244.597443185 & 22      & -2244.597443185 \\
      &UHF/cc-pVDZ    & 5 (14)  & -2244.708766895 & $> 500$ & -2244.6360      \\
      &U-B3LYP/cc-pVDZ& 9 (20)  & -2251.599470970 & 32      & -2251.599470969 \\
$^1$Fe$_4$S$_8$C$_4$H$_{12}^{2-}$\tnote{d}
      & RHF/cc-pVDZ   & 18 (44) & -8387.977176947 & $> 500$ & -8387.7108      \\
      & B3LYP/cc-pVDZ & 8 (21)  & -8399.568031935 & 165     & -8399.568031918 \\
    \hline
  \end{tabular}
\begin{tablenotes}\footnotesize
\item[a] Geometry is taken from Ref \cite{Daniels2000}
\item[b] Geometry is taken from Ref \cite{Hu2010}
\item[c] Geometry is taken from Ref \cite{Groenhof2005}
\item[d] Geometry is taken from Ref \cite{Sharma2014}
\item[e] Number in parenthesis is the total keyframes
\end{tablenotes}
\end{threeparttable}
\end{table*}

Table \ref{tab:alltests} presents the results of CIAH and DIIS for
some challenging SCF systems.\nocite{Sharma2014,Groenhof2005}
The CIAH algorithm shows better overall convergence than the DIIS algorithm.
Except $^3$UF$_4$ with U-LSDA/LANL2DZ, CIAH is able to converge all test
systems.
Using DIIS, 5 systems do not meet the convergence criteria within 500 SCF
iterations.
For the converged systems, there are 8 answers that CIAH and DIIS algorithms
show good agreements.
Aside from the 8 systems, CIAH and DIIS predicts closed solutions in 3 systems:
$^3$Cr$_2$ with U-LSDA/3-21G, $^1$UF$_4$ with B3LYP/LANL2DZ and $^3$UF$_4$ with
U-B3LYP/LANL2DZ.
In the 3 systems, CIAH solution is about 1 $mE_\text{h}$ (or less) lower than
DIIS solution.
For the rest systems (except $^3$UF$_4$ with U-LSDA which is not converged in
CIAH), noticeable differences can be found between the two algorithms:
the total energy predicted by CIAH algorithm is lower.
Most of these systems are associated to the unrestricted calculations.
These CIAH solutions have larger spin-contamination than that appeared in
DIIS solutions.
In these systems, spin-symmetry broken happens on the early
stage of CIAH iterations which is not observed in the entire DIIS iterations.
One possibility is the side effect of DIIS level shift.
Although level shift stabilizes the DIIS oscillations, it limits the
variational space that the optimization solver can reach.
Some DIIS solutions actually converge to the saddle point.
For example, feeding the DIIS solution of $^3$UF$_4$ with UHF/LANL2DZ
to the CIAH solver,
CIAH takes 7 extra iterations to move to the expected lower-energy answer
$E=-448.736476864$.

\subsection{Orbital localization}
\begin{figure}[htp]
  \begin{center}
    \includegraphics[width=0.5\textwidth]{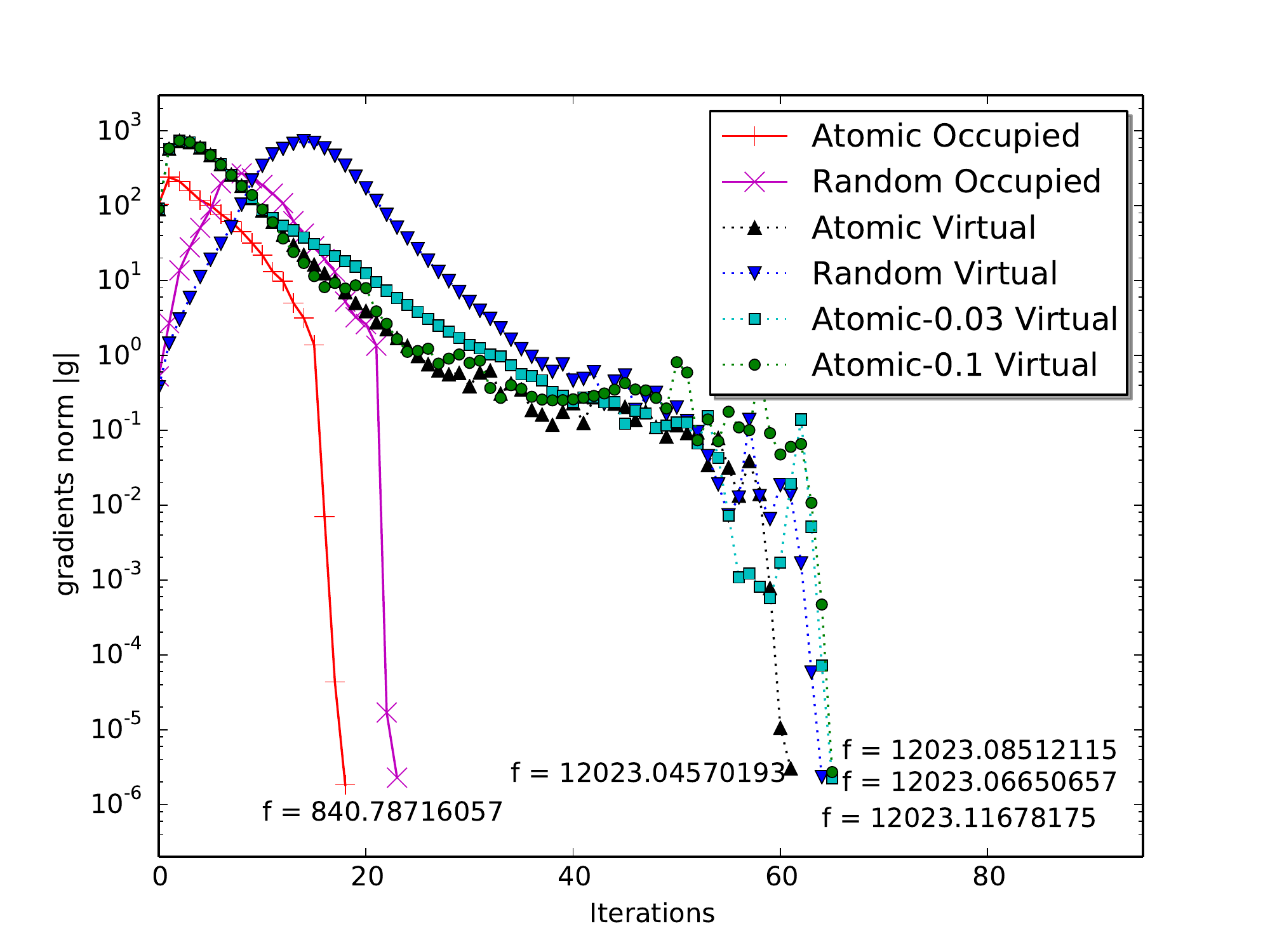}
  \end{center}
  \caption{Boys localization for C$_{60}$ molecule.
  The object function is $f = \sum_i \langle ii| (\mathbf{r}_1-\mathbf{r}_2)^2|ii\rangle$.}
  \label{fig:boys}
\end{figure}

CIAH algorithm can be applied with various type of orbital localization
methods.
Here we only demonstrated the convergence of Boys localization (see Figure
\ref{fig:boys}) for the HF occupied orbitals and virtual orbitals of buckyball
at RHF/cc-pVTZ level as an example.
Although not presented in this paper, Edmiston-Ruedenberg and Pipek-Mezey
localization can be accelerated by the same solver.

Unlike the SCF initial guess, orbital localization often starts with canonical
orbitals, which is typically very different to the final answer.
For buckyball, the canonical orbitals cannot be directly taken as
the initial guess for Boys localization because the orbital gradients are
strictly zero (at saddle point).
One solution is to add small noise on the initial guess to drive the
system out of the saddle point.  This is marked as ``random'' in Figure
\ref{fig:boys}.
Another initial guess we tried is pre-localization which is marked as
``atomic'' in Figure \ref{fig:boys}.
In the atomic initial guess, we compare the canonical orbitals $\psi$ with a
set of reference atomic orbitals $\chi$ to define the rotation $\tilde{U}$ in terms of the
SVD of the projection $\langle \chi|\psi\rangle $
\begin{gather*}
  \langle \chi|\psi\rangle = U\lambda V^\dagger, \\
  \tilde{U} = VU^\dagger.
\end{gather*}
Transformation $\tilde{U}$ thus defines the initial guess orbitals
$|\tilde{U}\psi\rangle$ which are close to the reference atomic orbitals.

For occupied orbitals, the two kinds of initial guess produce small difference
on the optimization procedure and converge to the same solution.
The random initial guess is slightly worse at beginning.
After about 7 steps to move out of the saddle point, it shows the similar
convergence curve as the atomic initial guess.
The orbital localization becomes difficult when diffused orbitals are involved.
There are many local minimum solutions close to each other.
For buckyball virtual orbitals, different initial guess (atomic and random),
different step size (0.03, 0.05, 0.1) lead to different solutions in which
the values of object function are differed by 0.05 au or around.
Regard to the system size and the total value of object function (12023 au), the
difference is negligible.

Since the initial guess is so different to the answer, one may expect that the
large step size is superior.
In fact, the tests for the three step size (0.03, 0.05 and 0.1) have closed
convergence performances.
They all take around 30 iterations to move away from the initial guess
and the next 30 iterations wandering around the quadratic region.
Once the solver steps into the quadratic region, they rapidly converge to the
solution.
Although not obvious in the figure, we observed during the optimization that
the step size 0.1 sometimes overshoots the displacement and causes
oscillation on the object function value.
This observation numerically supports that the small step size is able to
provide better gradients and Hessian estimations than the large step size.

\section{Conclusion}
In this work, we demonstrated a general second order algorithm CIAH for
orbital optimization problem.
In the CIAH algorithm, the optimization is embedded in the augmented
Hessian diagonalization procedure.
The evaluation of gradients and Hessian matrix are decoupled.
Various approximations can be used for the Hessian matrix.
By analyzing and comparing three Hessian approximations: density
fitting, sparse grids, basis projection, we find that the combination of
density fitting and sparse grids is able to produce high quality
approximations to the Hessian matrix.
Our numerical tests of SCF convergence suggests that CIAH is a stable,
reliable and efficient algorithm for SCF energy minimization problem.
Apart from the molecular SCF energy minimization,
CIAH algorithm can be used in many orbital optimization scenarios, such as
orbital localization, MCSCF orbital optimization, SCF energy minimization with
periodic boundary condition.
Our numerical tests of orbital localization verifies the capability of CIAH
algorithm to localize a large number of diffuse orbitals.

CIAH algorithm offers a new possible solution for a wide range of optimization
problem.
To make CIAH general purposed optimization algorithm, some problems remain for
further study.
First is the constraint optimization.
In the orbital optimization, the normalization constraints are imposed
indirectly by the exponential ansatz for the unitary transformation.  
Applying the normalization constraints in a direct manner can avoid the
expensive matrix exponential operation, which is particularly useful for large
scale systems.
Moreover, the constraint optimization would extend CIAH to the area of
geometry optimization and the transition state search.
Second is the step size for a general purpose optimization problem since the
small step assumption might not work efficiently in the general parameter space.

\section{Acknowledgments}
This work was supported by the National Science Foundation through NSF-CHE-1657286.

\bibliography{ref}

\begin{thebibliography}{25}%
\makeatletter
\providecommand \@ifxundefined [1]{%
 \@ifx{#1\undefined}
}%
\providecommand \@ifnum [1]{%
 \ifnum #1\expandafter \@firstoftwo
 \else \expandafter \@secondoftwo
 \fi
}%
\providecommand \@ifx [1]{%
 \ifx #1\expandafter \@firstoftwo
 \else \expandafter \@secondoftwo
 \fi
}%
\providecommand \natexlab [1]{#1}%
\providecommand \enquote  [1]{``#1''}%
\providecommand \bibnamefont  [1]{#1}%
\providecommand \bibfnamefont [1]{#1}%
\providecommand \citenamefont [1]{#1}%
\providecommand \href@noop [0]{\@secondoftwo}%
\providecommand \href [0]{\begingroup \@sanitize@url \@href}%
\providecommand \@href[1]{\@@startlink{#1}\@@href}%
\providecommand \@@href[1]{\endgroup#1\@@endlink}%
\providecommand \@sanitize@url [0]{\catcode `\\12\catcode `\$12\catcode
  `\&12\catcode `\#12\catcode `\^12\catcode `\_12\catcode `\%12\relax}%
\providecommand \@@startlink[1]{}%
\providecommand \@@endlink[0]{}%
\providecommand \url  [0]{\begingroup\@sanitize@url \@url }%
\providecommand \@url [1]{\endgroup\@href {#1}{\urlprefix }}%
\providecommand \urlprefix  [0]{URL }%
\providecommand \Eprint [0]{\href }%
\providecommand \doibase [0]{http://dx.doi.org/}%
\providecommand \selectlanguage [0]{\@gobble}%
\providecommand \bibinfo  [0]{\@secondoftwo}%
\providecommand \bibfield  [0]{\@secondoftwo}%
\providecommand \translation [1]{[#1]}%
\providecommand \BibitemOpen [0]{}%
\providecommand \bibitemStop [0]{}%
\providecommand \bibitemNoStop [0]{.\EOS\space}%
\providecommand \EOS [0]{\spacefactor3000\relax}%
\providecommand \BibitemShut  [1]{\csname bibitem#1\endcsname}%
\let\auto@bib@innerbib\@empty
\bibitem [{\citenamefont {Pulay}(1980)}]{Pulay1980}%
  \BibitemOpen
  \bibfield  {author} {\bibinfo {author} {\bibfnamefont {P.}~\bibnamefont
  {Pulay}},\ }\href {\doibase http://dx.doi.org/10.1016/0009-2614(80)80396-4}
  {\bibfield  {journal} {\bibinfo  {journal} {Chem. Phys. Lett.}\ }\textbf
  {\bibinfo {volume} {73}},\ \bibinfo {pages} {393 } (\bibinfo {year}
  {1980})}\BibitemShut {NoStop}%
\bibitem [{\citenamefont {Pulay}(1982)}]{Pulay1982}%
  \BibitemOpen
  \bibfield  {author} {\bibinfo {author} {\bibfnamefont {P.}~\bibnamefont
  {Pulay}},\ }\href {\doibase 10.1002/jcc.540030413} {\bibfield  {journal}
  {\bibinfo  {journal} {J. Comput. Chem.}\ }\textbf {\bibinfo {volume} {3}},\
  \bibinfo {pages} {556} (\bibinfo {year} {1982})}\BibitemShut {NoStop}%
\bibitem [{\citenamefont {Canc\`es}\ and\ \citenamefont
  {Le~Bris}(2000)}]{Cances2000}%
  \BibitemOpen
  \bibfield  {author} {\bibinfo {author} {\bibfnamefont {E.}~\bibnamefont
  {Canc\`es}}\ and\ \bibinfo {author} {\bibfnamefont {C.}~\bibnamefont
  {Le~Bris}},\ }\href {\doibase
  10.1002/1097-461X(2000)79:2<82::AID-QUA3>3.0.CO;2-I} {\bibfield  {journal}
  {\bibinfo  {journal} {Int. J. Quantum Chem.}\ }\textbf {\bibinfo {volume}
  {79}},\ \bibinfo {pages} {82} (\bibinfo {year} {2000})}\BibitemShut {NoStop}%
\bibitem [{\citenamefont {Saunders}\ and\ \citenamefont
  {Hillier}(1973)}]{Saunders1973}%
  \BibitemOpen
  \bibfield  {author} {\bibinfo {author} {\bibfnamefont {V.~R.}\ \bibnamefont
  {Saunders}}\ and\ \bibinfo {author} {\bibfnamefont {I.~H.}\ \bibnamefont
  {Hillier}},\ }\href {\doibase 10.1002/qua.560070407} {\bibfield  {journal}
  {\bibinfo  {journal} {Int. J. Quantum Chem.}\ }\textbf {\bibinfo {volume}
  {7}},\ \bibinfo {pages} {699} (\bibinfo {year} {1973})}\BibitemShut {NoStop}%
\bibitem [{\citenamefont {Kudin}, \citenamefont {Scuseria},\ and\ \citenamefont
  {Canc\`es}(2002)}]{Kudin2002}%
  \BibitemOpen
  \bibfield  {author} {\bibinfo {author} {\bibfnamefont {K.~N.}\ \bibnamefont
  {Kudin}}, \bibinfo {author} {\bibfnamefont {G.~E.}\ \bibnamefont {Scuseria}},
  \ and\ \bibinfo {author} {\bibfnamefont {E.}~\bibnamefont {Canc\`es}},\
  }\href {\doibase http://dx.doi.org/10.1063/1.1470195} {\bibfield  {journal}
  {\bibinfo  {journal} {J. Chem. Phys.}\ }\textbf {\bibinfo {volume} {116}},\
  \bibinfo {pages} {8255} (\bibinfo {year} {2002})}\BibitemShut {NoStop}%
\bibitem [{\citenamefont {Hu}\ and\ \citenamefont {Yang}(2010)}]{Hu2010}%
  \BibitemOpen
  \bibfield  {author} {\bibinfo {author} {\bibfnamefont {X.}~\bibnamefont
  {Hu}}\ and\ \bibinfo {author} {\bibfnamefont {W.}~\bibnamefont {Yang}},\
  }\href {\doibase http://dx.doi.org/10.1063/1.3304922} {\bibfield  {journal}
  {\bibinfo  {journal} {J. Chem. Phys.}\ }\textbf {\bibinfo {volume} {132}},\
  \bibinfo {eid} {054109} (\bibinfo {year} {2010}),\
  http://dx.doi.org/10.1063/1.3304922}\BibitemShut {NoStop}%
\bibitem [{\citenamefont {Canc\`es}(2001)}]{Cances2001}%
  \BibitemOpen
  \bibfield  {author} {\bibinfo {author} {\bibfnamefont {E.}~\bibnamefont
  {Canc\`es}},\ }\href {\doibase http://dx.doi.org/10.1063/1.1373430}
  {\bibfield  {journal} {\bibinfo  {journal} {J. Chem. Phys.}\ }\textbf
  {\bibinfo {volume} {114}},\ \bibinfo {pages} {10616} (\bibinfo {year}
  {2001})}\BibitemShut {NoStop}%
\bibitem [{\citenamefont {Th{\o}gersen}\ \emph {et~al.}(2004)\citenamefont
  {Th{\o}gersen}, \citenamefont {Olsen}, \citenamefont {Yeager}, \citenamefont
  {J{\o}rgensen}, \citenamefont {Sa{\l}ek},\ and\ \citenamefont
  {Helgaker}}]{Thoegersen2004}%
  \BibitemOpen
  \bibfield  {author} {\bibinfo {author} {\bibfnamefont {L.}~\bibnamefont
  {Th{\o}gersen}}, \bibinfo {author} {\bibfnamefont {J.}~\bibnamefont {Olsen}},
  \bibinfo {author} {\bibfnamefont {D.}~\bibnamefont {Yeager}}, \bibinfo
  {author} {\bibfnamefont {P.}~\bibnamefont {J{\o}rgensen}}, \bibinfo {author}
  {\bibfnamefont {P.}~\bibnamefont {Sa{\l}ek}}, \ and\ \bibinfo {author}
  {\bibfnamefont {T.}~\bibnamefont {Helgaker}},\ }\href {\doibase
  http://dx.doi.org/10.1063/1.1755673} {\bibfield  {journal} {\bibinfo
  {journal} {J. Chem. Phys.}\ }\textbf {\bibinfo {volume} {121}},\ \bibinfo
  {pages} {16} (\bibinfo {year} {2004})}\BibitemShut {NoStop}%
\bibitem [{\citenamefont {Th{\o}gersen}\ \emph {et~al.}(2005)\citenamefont
  {Th{\o}gersen}, \citenamefont {Olsen}, \citenamefont {K\"ohn}, \citenamefont
  {J{\o}rgensen}, \citenamefont {Sa{\l}ek},\ and\ \citenamefont
  {Helgaker}}]{Thoegersen2005}%
  \BibitemOpen
  \bibfield  {author} {\bibinfo {author} {\bibfnamefont {L.}~\bibnamefont
  {Th{\o}gersen}}, \bibinfo {author} {\bibfnamefont {J.}~\bibnamefont {Olsen}},
  \bibinfo {author} {\bibfnamefont {A.}~\bibnamefont {K\"ohn}}, \bibinfo
  {author} {\bibfnamefont {P.}~\bibnamefont {J{\o}rgensen}}, \bibinfo {author}
  {\bibfnamefont {P.}~\bibnamefont {Sa{\l}ek}}, \ and\ \bibinfo {author}
  {\bibfnamefont {T.}~\bibnamefont {Helgaker}},\ }\href {\doibase
  http://dx.doi.org/10.1063/1.1989311} {\bibfield  {journal} {\bibinfo
  {journal} {J. Chem. Phys.}\ }\textbf {\bibinfo {volume} {123}},\ \bibinfo
  {eid} {074103} (\bibinfo {year} {2005}),\
  http://dx.doi.org/10.1063/1.1989311}\BibitemShut {NoStop}%
\bibitem [{\citenamefont {H{\o}st}\ \emph {et~al.}(2008)\citenamefont
  {H{\o}st}, \citenamefont {Olsen}, \citenamefont {Jans\'ik}, \citenamefont
  {Th{\o}gersen}, \citenamefont {J{\o}rgensen},\ and\ \citenamefont
  {Helgaker}}]{Hoest2008}%
  \BibitemOpen
  \bibfield  {author} {\bibinfo {author} {\bibfnamefont {S.}~\bibnamefont
  {H{\o}st}}, \bibinfo {author} {\bibfnamefont {J.}~\bibnamefont {Olsen}},
  \bibinfo {author} {\bibfnamefont {B.}~\bibnamefont {Jans\'ik}}, \bibinfo
  {author} {\bibfnamefont {L.}~\bibnamefont {Th{\o}gersen}}, \bibinfo {author}
  {\bibfnamefont {P.}~\bibnamefont {J{\o}rgensen}}, \ and\ \bibinfo {author}
  {\bibfnamefont {T.}~\bibnamefont {Helgaker}},\ }\href {\doibase
  http://dx.doi.org/10.1063/1.2974099} {\bibfield  {journal} {\bibinfo
  {journal} {J. Chem. Phys.}\ }\textbf {\bibinfo {volume} {129}},\ \bibinfo
  {eid} {124106} (\bibinfo {year} {2008}),\
  http://dx.doi.org/10.1063/1.2974099}\BibitemShut {NoStop}%
\bibitem [{\citenamefont {Host}\ \emph {et~al.}(2008)\citenamefont {Host},
  \citenamefont {Jansik}, \citenamefont {Olsen}, \citenamefont {Jorgensen},
  \citenamefont {Reine},\ and\ \citenamefont {Helgaker}}]{Host2008}%
  \BibitemOpen
  \bibfield  {author} {\bibinfo {author} {\bibfnamefont {S.}~\bibnamefont
  {Host}}, \bibinfo {author} {\bibfnamefont {B.}~\bibnamefont {Jansik}},
  \bibinfo {author} {\bibfnamefont {J.}~\bibnamefont {Olsen}}, \bibinfo
  {author} {\bibfnamefont {P.}~\bibnamefont {Jorgensen}}, \bibinfo {author}
  {\bibfnamefont {S.}~\bibnamefont {Reine}}, \ and\ \bibinfo {author}
  {\bibfnamefont {T.}~\bibnamefont {Helgaker}},\ }\href {\doibase
  10.1039/B807639A} {\bibfield  {journal} {\bibinfo  {journal} {Phys. Chem.
  Chem. Phys.}\ }\textbf {\bibinfo {volume} {10}},\ \bibinfo {pages} {5344}
  (\bibinfo {year} {2008})}\BibitemShut {NoStop}%
\bibitem [{\citenamefont {Wang}\ \emph {et~al.}(2011)\citenamefont {Wang},
  \citenamefont {Yam}, \citenamefont {Chen},\ and\ \citenamefont
  {Chen}}]{Wang2011}%
  \BibitemOpen
  \bibfield  {author} {\bibinfo {author} {\bibfnamefont {Y.~A.}\ \bibnamefont
  {Wang}}, \bibinfo {author} {\bibfnamefont {C.~Y.}\ \bibnamefont {Yam}},
  \bibinfo {author} {\bibfnamefont {Y.~K.}\ \bibnamefont {Chen}}, \ and\
  \bibinfo {author} {\bibfnamefont {G.}~\bibnamefont {Chen}},\ }\href {\doibase
  http://dx.doi.org/10.1063/1.3609242} {\bibfield  {journal} {\bibinfo
  {journal} {J. Chem. Phys.}\ }\textbf {\bibinfo {volume} {134}},\ \bibinfo
  {eid} {241103} (\bibinfo {year} {2011}),\
  http://dx.doi.org/10.1063/1.3609242}\BibitemShut {NoStop}%
\bibitem [{\citenamefont {Chen}\ and\ \citenamefont {Wang}(2011)}]{Chen2011}%
  \BibitemOpen
  \bibfield  {author} {\bibinfo {author} {\bibfnamefont {Y.~K.}\ \bibnamefont
  {Chen}}\ and\ \bibinfo {author} {\bibfnamefont {Y.~A.}\ \bibnamefont
  {Wang}},\ }\href {\doibase 10.1021/ct2004512} {\bibfield  {journal} {\bibinfo
   {journal} {J. Chem. Theory Comput.}\ }\textbf {\bibinfo {volume} {7}},\
  \bibinfo {pages} {3045} (\bibinfo {year} {2011})}\BibitemShut {NoStop}%
\bibitem [{\citenamefont {Daniels}\ and\ \citenamefont
  {Scuseria}(2000)}]{Daniels2000}%
  \BibitemOpen
  \bibfield  {author} {\bibinfo {author} {\bibfnamefont {A.~D.}\ \bibnamefont
  {Daniels}}\ and\ \bibinfo {author} {\bibfnamefont {G.~E.}\ \bibnamefont
  {Scuseria}},\ }\href {\doibase 10.1039/B000618L} {\bibfield  {journal}
  {\bibinfo  {journal} {Phys. Chem. Chem. Phys.}\ }\textbf {\bibinfo {volume}
  {2}},\ \bibinfo {pages} {2173} (\bibinfo {year} {2000})}\BibitemShut
  {NoStop}%
\bibitem [{\citenamefont {Fischer}\ and\ \citenamefont
  {Alml\"of}(1992)}]{Fischer1992}%
  \BibitemOpen
  \bibfield  {author} {\bibinfo {author} {\bibfnamefont {T.~H.}\ \bibnamefont
  {Fischer}}\ and\ \bibinfo {author} {\bibfnamefont {J.}~\bibnamefont
  {Alml\"of}},\ }\href {\doibase 10.1021/j100203a036} {\bibfield  {journal}
  {\bibinfo  {journal} {J. Phys. Chem.}\ }\textbf {\bibinfo {volume} {96}},\
  \bibinfo {pages} {9768} (\bibinfo {year} {1992})}\BibitemShut {NoStop}%
\bibitem [{\citenamefont {Neese}(2000)}]{Neese2000}%
  \BibitemOpen
  \bibfield  {author} {\bibinfo {author} {\bibfnamefont {F.}~\bibnamefont
  {Neese}},\ }\href {\doibase http://dx.doi.org/10.1016/S0009-2614(00)00662-X}
  {\bibfield  {journal} {\bibinfo  {journal} {Chem. Phys. Lett.}\ }\textbf
  {\bibinfo {volume} {325}},\ \bibinfo {pages} {93 } (\bibinfo {year}
  {2000})}\BibitemShut {NoStop}%
\bibitem [{\citenamefont {Lengsfield}(1980)}]{Lengsfield1980}%
  \BibitemOpen
  \bibfield  {author} {\bibinfo {author} {\bibfnamefont {B.~H.}\ \bibnamefont
  {Lengsfield}},\ }\href {\doibase 10.1063/1.439885} {\bibfield  {journal}
  {\bibinfo  {journal} {J. Chem. Phys.}\ }\textbf {\bibinfo {volume} {73}},\
  \bibinfo {pages} {382} (\bibinfo {year} {1980})}\BibitemShut {NoStop}%
\bibitem [{\citenamefont {J{\o}rgensen}, \citenamefont {Swanstrom},\ and\
  \citenamefont {Yeager}(1983)}]{Jorgensen1983}%
  \BibitemOpen
  \bibfield  {author} {\bibinfo {author} {\bibfnamefont {P.}~\bibnamefont
  {J{\o}rgensen}}, \bibinfo {author} {\bibfnamefont {P.}~\bibnamefont
  {Swanstrom}}, \ and\ \bibinfo {author} {\bibfnamefont {D.~L.}\ \bibnamefont
  {Yeager}},\ }\href {\doibase http://dx.doi.org/10.1063/1.444508} {\bibfield
  {journal} {\bibinfo  {journal} {J. Chem. Phys.}\ }\textbf {\bibinfo {volume}
  {78}},\ \bibinfo {pages} {347} (\bibinfo {year} {1983})}\BibitemShut
  {NoStop}%
\bibitem [{\citenamefont {Jensen}\ and\ \citenamefont
  {J{\o}rgensen}(1984)}]{Jensen1984}%
  \BibitemOpen
  \bibfield  {author} {\bibinfo {author} {\bibfnamefont {H.-J.~A.}\
  \bibnamefont {Jensen}}\ and\ \bibinfo {author} {\bibfnamefont
  {P.}~\bibnamefont {J{\o}rgensen}},\ }\href {\doibase
  http://dx.doi.org/10.1063/1.446797} {\bibfield  {journal} {\bibinfo
  {journal} {J. Chem. Phys.}\ }\textbf {\bibinfo {volume} {80}},\ \bibinfo
  {pages} {1204} (\bibinfo {year} {1984})}\BibitemShut {NoStop}%
\bibitem [{\citenamefont {Sun}\ and\ \citenamefont {Chan}(2017)}]{CASSCF}%
  \BibitemOpen
  \bibfield  {author} {\bibinfo {author} {\bibfnamefont {Q.}~\bibnamefont
  {Sun}}\ and\ \bibinfo {author} {\bibfnamefont {G.~K.-L.}\ \bibnamefont
  {Chan}},\ }\href@noop {} {\bibfield  {journal} {\bibinfo  {journal} {Chem.
  Phys. Lett.}\ } (\bibinfo {year} {2017})},\ \bibinfo {note}
  {submitted}\BibitemShut {NoStop}%
\bibitem [{\citenamefont {Ghosh}\ \emph {et~al.}(2008)\citenamefont {Ghosh},
  \citenamefont {Hachmann}, \citenamefont {Yanai},\ and\ \citenamefont
  {Chan}}]{Ghosh2008}%
  \BibitemOpen
  \bibfield  {author} {\bibinfo {author} {\bibfnamefont {D.}~\bibnamefont
  {Ghosh}}, \bibinfo {author} {\bibfnamefont {J.}~\bibnamefont {Hachmann}},
  \bibinfo {author} {\bibfnamefont {T.}~\bibnamefont {Yanai}}, \ and\ \bibinfo
  {author} {\bibfnamefont {G.~K.-L.}\ \bibnamefont {Chan}},\ }\href {\doibase
  http://dx.doi.org/10.1063/1.2883976} {\bibfield  {journal} {\bibinfo
  {journal} {J. Chem. Phys.}\ }\textbf {\bibinfo {volume} {128}},\ \bibinfo
  {eid} {144117} (\bibinfo {year} {2008}),\
  http://dx.doi.org/10.1063/1.2883976}\BibitemShut {NoStop}%
\bibitem [{\citenamefont {Davidson}(1975)}]{Davidson1975}%
  \BibitemOpen
  \bibfield  {author} {\bibinfo {author} {\bibfnamefont {E.~R.}\ \bibnamefont
  {Davidson}},\ }\href {\doibase 10.1016/0021-9991(75)90065-0} {\bibfield
  {journal} {\bibinfo  {journal} {J. Comput. Phys.}\ }\textbf {\bibinfo
  {volume} {17}},\ \bibinfo {pages} {87 } (\bibinfo {year} {1975})}\BibitemShut
  {NoStop}%
\bibitem [{\citenamefont {Sun}(2014)}]{PYSCF}%
  \BibitemOpen
  \bibfield  {author} {\bibinfo {author} {\bibfnamefont {Q.}~\bibnamefont
  {Sun}},\ }\href@noop {} {\enquote {\bibinfo {title} {Python module for
  quantum chemistry program},}\ }\bibinfo {howpublished}
  {\url{https://github.com/sunqm/pyscf.git}} (\bibinfo {year}
  {2014})\BibitemShut {NoStop}%
\bibitem [{\citenamefont {Groenhof}\ \emph {et~al.}(2005)\citenamefont
  {Groenhof}, \citenamefont {Swart}, \citenamefont {Ehlers},\ and\
  \citenamefont {Lammertsma}}]{Groenhof2005}%
  \BibitemOpen
  \bibfield  {author} {\bibinfo {author} {\bibfnamefont {A.~R.}\ \bibnamefont
  {Groenhof}}, \bibinfo {author} {\bibfnamefont {M.}~\bibnamefont {Swart}},
  \bibinfo {author} {\bibfnamefont {A.~W.}\ \bibnamefont {Ehlers}}, \ and\
  \bibinfo {author} {\bibfnamefont {K.}~\bibnamefont {Lammertsma}},\ }\href
  {\doibase 10.1021/jp0441442} {\bibfield  {journal} {\bibinfo  {journal} {J.
  Phys. Chem. A}\ }\textbf {\bibinfo {volume} {109}},\ \bibinfo {pages} {3411}
  (\bibinfo {year} {2005})}\BibitemShut {NoStop}%
\bibitem [{\citenamefont {Sharma}\ \emph {et~al.}(2014)\citenamefont {Sharma},
  \citenamefont {Sivalingam}, \citenamefont {Neese},\ and\ \citenamefont
  {Kin-Lic}}]{Sharma2014}%
  \BibitemOpen
  \bibfield  {author} {\bibinfo {author} {\bibfnamefont {S.}~\bibnamefont
  {Sharma}}, \bibinfo {author} {\bibfnamefont {K.}~\bibnamefont {Sivalingam}},
  \bibinfo {author} {\bibfnamefont {F.}~\bibnamefont {Neese}}, \ and\ \bibinfo
  {author} {\bibfnamefont {C.}~\bibnamefont {Kin-Lic}},\ }\href
  {http://dx.doi.org/10.1038/nchem.2041} {\bibfield  {journal} {\bibinfo
  {journal} {Nat. Chem.}\ }\textbf {\bibinfo {volume} {6}},\ \bibinfo {pages}
  {927} (\bibinfo {year} {2014})}\BibitemShut {NoStop}%
\end{thebibliography}%

\end{document}